\documentstyle[11pt,newpasp,twoside]{article}
\def\edcomment#1{\iffalse\marginpar{\raggedright\sl#1\/}\else\relax\fi}
\reversemarginpar

\begin{document}
\title{Tracing Cosmic Evolution with an {\em XMM} Serendipitous Cluster Survey}
 \author{A.~Kathy Romer}
\affil{Physics Department, Carnegie Mellon University, 5000
Forbes Avenue, Pittsburgh, PA 15213, USA}
\author{Pedro T.~P.~Viana}
\affil{Centro de Astrof\'{\i}sica da Universidade do Porto,
Rua das Estrelas s/n, 4150 Porto, Portugal}

\author{Christopher A.~Collins}\affil{Astrophysics Research Institute,
Liverpool John Moores University, U.K.}

\author{Andrew R.~Liddle} \affil{ Astronomy
Centre, University of Sussex, Brighton, UK.}

\author{Robert G.~Mann}  \affil{Institute for Astronomy, University of
Edinburgh, Royal Observatory, Blackford Hill, Edinburgh, UK.}

\author{Robert C.~Nichol}\affil{Carnegie Mellon University, Physics
Department, 5000 Forbes Avenue, Pittsburgh, PA 15213, USA.}

\begin{abstract}

This paper describes updated predictions, as a function of the
underlying cosmological model, for a serendipitous galaxy cluster
survey that we plan to conduct with the {\em XMM-Newton} X-ray
Satellite.  We have included the effects of the higher than
anticipated internal background count rates and have expanded our
predictions to include clusters detected at $>3\sigma$. Even with the
enhanced background levels, we expect the XCS to detect sufficient
clusters at $z>1$ to differentiate between open and flat cosmological
models. We have compared the XCS cluster redshift distribution to
those expected from the {\em XMM} Slew Survey and the {\em ROSAT}
Massive Cluster Survey (MACS) and find them to be complementary. We
conclude that the future existence of the XCS should not deter the
launch of a dedicated X-ray survey satellite.

\end{abstract}

\section{Introduction}

In an earlier paper (Romer et al. 2001, R01 hereafter) we described
the expected catalog properties and scientific applications of an {\em
XMM-Newton} Cluster Survey (XCS) based on serendipitous detections in
pointed observations.  We revisit these predictions below, in light of
improved knowledge of the in flight performance of {\em XMM}. We also
compare the XCS to other on-going X-ray surveys.

\begin{table}[t]

\label{cluster-numbers}
\begin{center}

\begin{tabular}{|c|ccc|ccc|ccc|}\hline
& 
\multicolumn{3}{|c|}{\bf $\Omega_0=0.3, \Omega_\Lambda=0.7$} 
&\multicolumn{3}{|c|}{\bf $\Omega_0=1.0, \Omega_\Lambda=0.0$} 
&\multicolumn{3}{|c|}{\bf $\Omega_0=0.3, \Omega_\Lambda=0.0$}\\
$T>4$ keV        
& $>8\sigma$ & $>8\sigma$ & $>3\sigma$
& $>8\sigma$ & $>8\sigma$ & $>3\sigma$
& $>8\sigma$ & $>8\sigma$ & $>3\sigma$\\
&$i$         & $ii$       & $iii$
&$i$         & $ii$       & $iii$
&$i$         & $ii$       & $iii$ \\ \hline
$z>0$   & 750 & 720 & 890 &  80 & 80 & 80&  1100 & 1010 & 1530\\
$z>0.3$ & 700 & 660 & 830 &  50 & 50 & 50&  1060$^{\ast}$ & 970  & 1490 \\ 
$z>1$   & 170 & 150 & 300 &   1 &  1 &  2&   480 & 390  & 900 \\ \hline
\end{tabular}
\end{center}

\caption{The expected number of $T>4$ keV clusters detected by the XCS
as a function of cosmology. Column $i$, the original XCS predictions
for $>8\sigma$ detections from R01. Columns $ii$ and $iii$, updated
XCS predictions based on the measured in flight internal background
count rate. All the results assume a final survey area of 800
sq. degrees. $^{\ast}$There was a typo in the the printed version of
R01, this is the correct value.}
\end{table}

\section {The Effect of an Enhanced Internal Background Count Rate on
the XCS}

Recently it has come to light that the quiescent internal background
count rates in the EPIC detectors are roughly ten times higher than
was anticipated prior to the {\em XMM-Newton} launch (see Lumb
2001). We have, therefore, recomputed the expected properties of the
XCS cluster catalog using a total (external plus internal) background
level that is roughly double that used in R01.  In Table~1, we compare
the new predictions (column {\em ii}) with those published in Table~4
of R01 (column {\em i}).  For example, for an open model
($\Omega_0=0.3$, $\Omega_{\Lambda}=0$), we now predict that the XCS
will detect 390 $T>4$ keV, $z>1$ clusters at $>8 \sigma$, compared to
480 in R01. Changes at that level should not severely limit the
ability of XCS to differentiate between open and flat cosmological
models.

We have also included in Table~1 predictions for the number of XCS
clusters that would be detected at $>3\sigma$. We did not include low
signal to noise detections in R01 because they will be hard to
identify using an extent criterion alone. However, when combined with
Planck and/or SDSS data, it may be possible to identify these objects
in a  timely and quantifiable manner, so we include them here for
completeness. 

The redshift distribution of the $>3 \sigma$ and $>8 \sigma$ XCS
detections can be seen in Figure~1.

\section {Comparisons with the MACS and {\em XMM} Slew Surveys}

\begin{figure}[t]
\plotfiddle{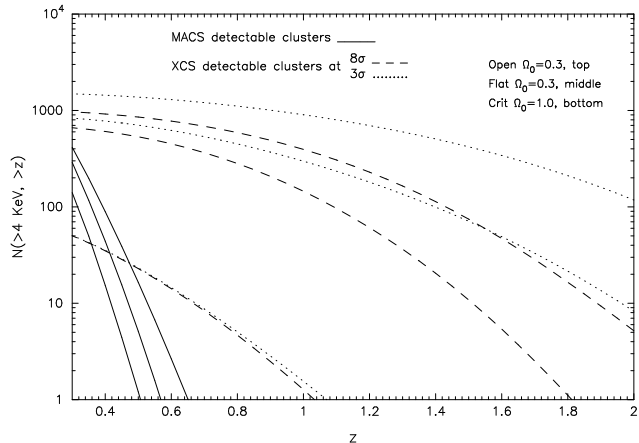}{2in}{270}{35}{35}{-150}{175}

\caption{Predicted redshift distribution ($z>0.3$) of MACS clusters
(solid lines) and of the XCS Survey (dashed and dotted lines).}

\end{figure}

 The Massive Cluster Survey (MACS, Ebeling, Edge \& Henry 2001) has
been very successful at identifying high redshift, high luminosity,
clusters detected in 22,735 square degrees of the ROSAT All Sky
Survey. The wide areal coverage of the MACS (nearly 30 times that of
the XCS!) results in sensitivity to a very interesting class of
clusters, those at medium redshift and high luminosity
($L_x>5\times10^{44}$ erg~cm$^{-2}$~s$^{-1}$; 0.1-2.0 keV). MACS
clusters are already being used for a variety science programs including
studies of cluster evolution and the Sunyaev-Zel'dovich Effect.

Using the R01 methodology and based on the MACS selection function
presented in Figure~5 of Ebeling et al. (2001) we have been able to
predict the number of clusters that would be detected by MACS as a
function of cosmology. We present those results in Figure~1 and
Table~1. Overall, we can expect MACS to detect between 100 and 500
clusters at $z>0.3$, depending on cosmology.

\begin{figure}
\plotfiddle{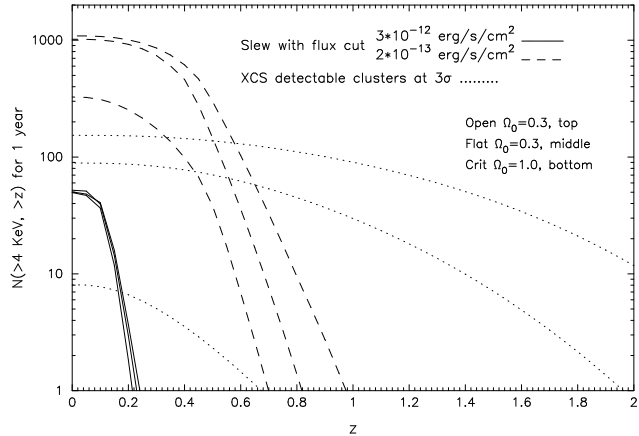}{2in}{270}{35}{35}{-150}{175}

\caption{Predicted redshift distribution of the clusters detected
during the first year of an {\em XMM} Slew Survey (solid and dashed
lines) and of the XCS Survey (dotted lines).}
\end{figure}
%shouldn't say ``first year''

An  {\em XMM} Slew Survey has been proposed ({\it e.g.}  Lumb \& Jones
2000). An {\em XMM} Slew survey would have two important advantages
over the XCS; it would be drawn from random parts of the sky (rather
than from regions that surround known X-ray targets) and it would
cover significantly more area. Lumb \& Jones (2000) estimated an
annual survey rate of 4000 square degrees to a flux limit of
$2\times10^{-13}$ erg~cm$^{-2}$~s$^{-1}$ (0.5-2.0 keV). This compares
to an estimated annual survey rate for the XCS of $\simeq 80$ square
degrees (the XCS catalog will not have a single flux limit, but a typical
value will be $\simeq 1.5\times10^{-14}$ erg~cm$^{-2}$~s$^{-1}$ in the
0.5-2.0 keV band, see R01). 

We have attempted to predict the properties of a Slew Survey cluster
catalog using the R01 methodology. At the time of writing, the Slew
Survey had only just begun and the actual sensitivity level was
unknown (Lumb, private communication). We, therefore, present results
for two different (0.5-2.0 keV) flux limits; $2\times10^{-13}$
erg~cm$^{-2}$~s$^{-1}$ and $3\times10^{-12}$ erg~cm$^{-2}$~s$^{-1}$,
see Figure~2. We have assumed that 4000 square degrees would be
covered in one year by the Slew Survey.  For comparison we have shown
the expected numbers of clusters that would be detected at $>3 \sigma$
in one year of the XCS (or 80 square degrees).

A Slew Survey with $2\times10^{-13}$ erg~cm$^{-2}$~s$^{-1}$
sensitivity would produce a very impressive cluster catalog at low and
medium redshifts. Such a catalog would have a variety of applications
including constraints on $\sigma_8$ and $\Omega_0$ and studies of
cluster evolution and Planck foregrounds.

In summary, the MACS and the {\em XMM} Slew Surveys have much better
low redshift sensitivity than XCS by virtue of their large areal
coverage. However, neither will be able to differentiate between open
and flat low $\Omega_{0}$ models, because they lack sensitivity to
high redshift clusters.  These surveys should be seen as complementary
too, not in competition with, the XCS.

The XCS faces many challenges, including the complexity of its
selection function and our lack of understanding of cluster
evolution. The future existence of the XCS should not deter the launch
of a dedicated X-ray survey satellite.

We would like to acknowledge the assistance of Juliet McKensie, Shane Zabel and Kevin Bandura with the talk preparations.

%did not mention temperature measurements or area surveyed to date

\end{document}